\documentclass[preprints]{JHEP3}
\usepackage{epsfig,graphicx}

\title{Genuine Symmetry of Staggered Fermion}

\author{Katsumi Itoh\\Department of Education, Niigata University, Ikarashi 2-8050, Niigata 950-2181, Japan\\
	E-mail: \email{itoh@ed.niigata-u.ac.jp}}
\author{Mitsuhiro Kato\\Institute of Physics,~University of
	Tokyo, Komaba, Meguroku, Tokyo 153-8902, Japan\\
	E-mail: \email{kato@hep1.c.u-tokyo.ac.jp}}
\author{Michika Murata,~Hideyuki Sawanaka\\
Graduate School of Science and Technology,~Niigata University,~Ikarashi 2-8050, Niigata 950-2181, Japan\\
	E-mail: \email{michika@muse.sc.niigata-u.ac.jp},~\email{hide@muse.sc.niigata-u.ac.jp}}
\author{Hiroto So\\Department of Physics,~Niigata University, Ikarashi 2-8050, Niigata 950-2181, Japan\\
	E-mail: \email{so@muse.sc.niigata-u.ac.jp}}

\preprint{NIIG-DP-04-3\\UT-Komaba/04-14}	

\abstract{We present a new formulation of the staggered fermion
on the $D$-dimensional lattice based on the $SO(2D)$ Clifford algebra,
which is naturally present in the action.  The action of the massless
staggered fermion is invariant under the discrete rotation and the
$SO(2D)$ chiral and other discrete transformations.  From transformation
properties of the fermion, we find two local meson operators (one scalar
and one pseudoscalar) in addition to two standard meson operators.}

\keywords{lgf, sts}

\begin{document}

\section{Introduction}    

The fermions on lattice are bound to suffer from the doubling problem.
In the early stage of the development of the lattice gauge theory, it
is proposed that massless modes due to the doubling may be
regarded as internal (flavor) degrees of freedom
\cite{Kogut:1974ag,Susskind:1976jm}.  The naively discretized Dirac
fermion was formulated as a staggered fermion in
Ref.~\cite{Kawamoto:1981hw}.  The reconstruction procedure was found to
give spinors with flavor degrees of freedom for the staggered
fermion with one component spinor on each
site~\cite{Kluberg-Stern:1983dg}.

Since the spinor and flavor degrees of freedom originate from its
geometrical structure, the staggered fermion must transform nontrivially
under the rotation.  To clarify it is one of the main motivations of the
present paper.  In this paper, we consider the one component staggered
fermion in $D$ dimensions, which has $2^{D}$ degrees of freedom due to
the doubling.  The number coincides with the dimension of the spinor
representation of $SO(2D)$.  Indeed, we will show that the $SO(2D)$
Clifford algebra is naturally present in the theory and it plays a
crucial role to study symmetries of the staggered Dirac operator.
We study the (discrete) rotational symmetry, the chiral symmetry and
other discrete transformations such as parity, charge conjugation and
time reversal.

Symmetries of the staggered fermion have been discussed earlier in
Ref.\cite{Doel-Smit:1983dg}.  Our approach differs from theirs on two
important points: 1) The $SO(2D)$ structure is fully respected ; 2) We
use a different definition for the rotation, which, we believe,
is more suitable for the staggered fermion.  The details will be
discussed in section 3.

With the knowledge of symmetries, we find two extra meson operators,
more than those well-known scalar and pseudoscalar operators.  These
operators have been overlooked in simulations and it is very important
to take them into account for a reliable study.

\section{The Staggered Fermion and $SO(2D)$ Clifford Algebra}

In this section, we present a new expression of the staggered Dirac
operator based on the $SO(2D)$ Clifford algebra to be found for the
$D$-dimensional lattice.  For the purpose, it is important to understand
the geometrical structure associated with the staggered fermion and its
relation to the $SO(2D)$ Clifford algebra.  This will be explained in the
next subsection.  Then we proceed to the expression of the staggered
Dirac operator that respects the algebra.

\subsection{Geometrical structure of the staggered fermion}

To clearly state the geometrical structure, we classify links,
plaquettes and hypercubes.\footnote{Some notions presented here are
refined versions of those reported
earlier \cite{Itoh:2001rx,Itoh:2002nq}.}  

Consider the link directed to
the positive $\mu$ direction from the site $n_{\mu}$.  We denote it by
$(n, \mu)$ and assign the sign factor, $(-1)^{n_{\mu}}$.  As for the
link directed to the negative direction, we write it as $(n, -\mu)$ and 
the associated sign factor must be $(-1)^{n_{\mu}+1}$.
The plaquette $(n,\mu\nu)$
is defined as the ordered links,
\begin{equation}
(n,\mu\nu) \equiv
 (n,\mu)(n+\hat{\mu},\nu)(n+\hat{\mu}+\hat{\nu},-\mu)(n+\hat{\nu},-\nu)~~~~~(\mu<\nu).
\label{plaquette}
\end{equation}
On each link of $(n, \mu \nu)$, we have a sign factor.  We refer to the
plaquette with $(++++)$ or $(----)$ sign factors as a cell-plaquette,
while that with a set of mixed signs as a pipe-plaquette.  On any two
dimensional surface of the $D$-dimensional lattice, we find the
checkered pattern, or the Ichimatsu pattern, formed by the cell- and
pipe-plaquettes.

Note that there are distinctive hypercubes on the lattice: Those formed
solely by cell-plaquettes with all plus (minus) sign factors will be
called microcells (macrocells) by the reason to be explained shortly.
The coordinates $(n_{\mu})$ of sites on a microcell can be written as
$n_{\mu}=2N_{\mu}+r_{\mu}$, where $N_{\mu}$ is some integer and
$r_{\mu}$ takes the value of $0$ or $1$.  As for a macrocell, those
are written similarly as $n_{\mu}=2N_{\mu}+1+r_{\mu}$.  

\FIGURE[pos]{\epsfig{file=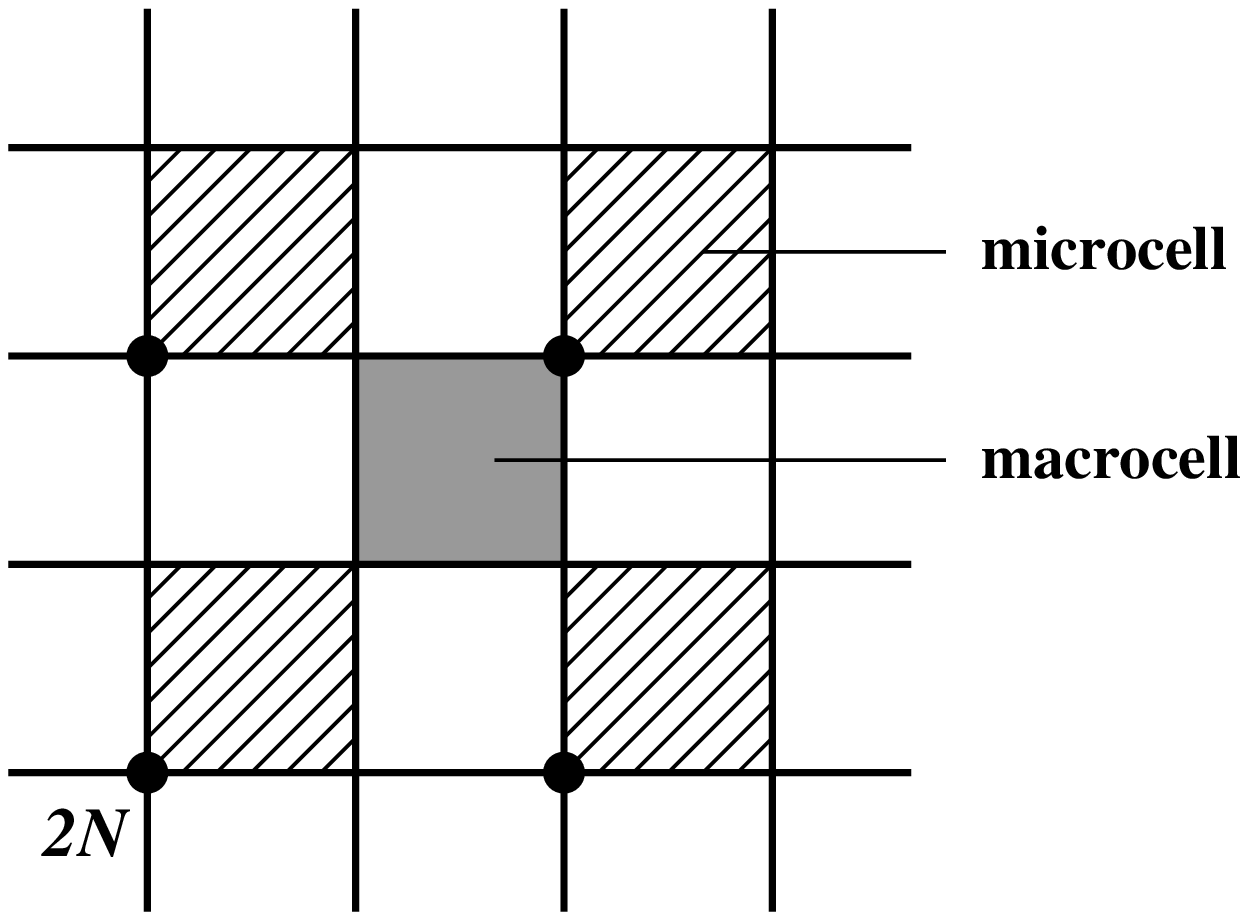,width=7.7cm}\caption{The microcells and macrocells}}

After having introduced the notions of micro- and macrocells, we observe
a slightly more detailed structure on a two dimensional surface than the
Ichimatsu pattern.  The $D=2$ example is shown in Fig.1, where a
microcell (macrocell) is shown as a shaded (gray) square.  We would find
the same pattern on any two dimensional surface of the $D$-dimensional
lattice.  It is noteworthy to realize that the pattern is invariant
under the translation by twice the lattice spacing, $2a$.  We call this
translation the modulo 2 translation.

After the reconstruction, the Grassmann variables on the sites of a
microcell are to form a spinor labelled by $(N_{\mu})$\footnote{In this
sense, $(N_{\mu})$ can be regarded as the coordinate of a microcell.}.
The relative coordinates $(r_{\mu})$ on the Grassmann variables are
transformed into spinor and flavor indices.  In this sense, a microcell
may be regarded as an internal space.  The global structure such as the
fermion kinetic term are formed out of the remaining geometrical
structure.

The relation of the staggered fermion to the $SO(2D)$ Clifford algebra
comes from the fact that sites on a microcell can be regarded as the
weight lattice for the spinor representation of $SO(2D)$.  Therefore the
Grassmann variables on sites in a microcell are in the spinor
representation.  Accordingly, we will find that the Dirac operator has a
natural expression in terms of gamma matrices associated with this
$SO(2D)$.

In the next section, we will consider the symmetry of the action under
the $\pi/2$ rotation around the center of a microcell.  One could have
considered the $\pi/2$ rotation around a site.  Actually, the latter
rotation was studied earlier in Ref.~\cite{Doel-Smit:1983dg} and the
rotational symmetry of the staggered fermion was established.  These two
rotations are to end up with the same rotational symmetry in the
continuum limit.  We choose the former definition to respect the
geometrical structure explained above that is inherent to the staggered
fermion.\footnote{The $\pi /2$ rotations around a center of any
hypercube other than a microcell also leave the geometrical structure
invariant.  However, we do not further consider these possibilities for
the rotational symmetry in this paper.}

\subsection{The staggered Dirac operator} 

The single component staggered Dirac operator is given as
\begin{eqnarray}
(D_{\rm st})_{n,n'} & = & \sum_{\mu} \eta_{\mu}(n)
\frac{\delta_{n',n+\hat{\mu}} U^{\rm rep}_{n,\mu} -  \delta_{n',n-\hat{\mu}} 
U^{{\rm rep} \hspace{0.5mm} \dagger}_{n-\hat{\mu},\mu}}{2a},  
\label{staggered Dirac op}
\end{eqnarray}
where $\eta_{\mu}(n) = \eta_{\mu}(r) = (-1)^{\sum_{\nu <\mu}r_{\nu}}$.
The link variable $U^{\rm rep}_{n,\mu}$ acts on the fermion
variable in the representation ``rep" of the gauge group.  
For simplicity, we drop the superscript ``rep" from the
link variable in the rest of the paper.

In preparation to rewrite the operator in eq.~(\ref{staggered Dirac op})
to act on the coordinates $(r,N)$, we introduce notations for gamma
matrices and their products for the $SO(2D)$ Clifford algebra.  By
$\vec{\varepsilon}$, we denote a bit-valued $D$-dimensional vector: That
is, each entry takes the value of $0$ or $1$. For each
$\vec{\varepsilon}$, we define the matrix
\newpage
\begin{eqnarray}
(\Gamma_{\mu,\vec{\varepsilon}})_{(r_1,r'_1)\cdots (r_D,r'_D)}
& \equiv&\Bigl((\sigma_3{}^{\varepsilon_1}
\otimes\cdots\otimes\sigma_3{}^{\varepsilon_D})
\label{generic Gamma}\\
&{}&\times
(\sigma_3\otimes\cdots\otimes\sigma_3\otimes
\stackrel{\raise5pt\hbox{\scriptsize$\mu$}\mbox{ }}{\sigma_1}
\otimes{\bf 1}\otimes\cdots\otimes{\bf 1})\Bigr)_{(r_1,r'_1)\cdots
(r_D,r'_D)}. 
\nonumber
\end{eqnarray}
As special cases, we obtain a basis for the $SO(2D)$ Clifford algebra,
\begin{equation}
\gamma_{\mu} = \Gamma_{\mu,\vec{0}}, ~~~~~  \tilde{\gamma}_{\mu} = 
-i\Gamma_{\mu,\vec{e}_{\mu}},
\label{SO(2D) basis}
\end{equation}
with $\vec{e}_{\mu}$ being the unit vector along the $\mu$-direction.
Note that $\gamma_{\mu}$ in eq.~(\ref{SO(2D) basis}) are for an
irreducible representation of the $SO(2D)$ Clifford algebra while they are
reducible as for the $SO(D)$ Clifford algebra.  In terms of the basis
given in eq.~(\ref{SO(2D) basis}), eq.~(\ref{generic Gamma}) may be
expressed as
\begin{eqnarray}
(\Gamma_{\mu,\vec{\varepsilon}})_{(r_1,r'_1)\cdots (r_D,r'_D)}
=\Bigl((i\tilde{\gamma}_1 {\gamma_1})^{\varepsilon_1} \cdots (i\tilde{\gamma}_D 
{\gamma_D})^{\varepsilon_D} \gamma_{\mu}\Bigr)_{(r_1,r'_1)\cdots (r_D,r'_D)}  .
\end{eqnarray}

We may rewrite $(D_{\rm st})_{n,n'}$ as an operator acting on the coordinates $(r,N)$.
That is realized by using the expression (\ref{generic Gamma}),
\begin{eqnarray}
(D_{\rm st})_{n,n'} \equiv (D_{\rm st})_{(r,N),(r',N')}  
&=& \sum_{\mu,\vec{\varepsilon}} 
(\Gamma_{\mu,\vec{\varepsilon}})_{(r,r')} 
(D_{\mu,\vec{\varepsilon}})_{(N,N')}.
\label{(r,N) Dirac op}
\end{eqnarray}
Here $D_{\mu,\vec{\varepsilon}}$ is the generalized difference operator, 
\begin{eqnarray}
(D_{\mu,\vec{\varepsilon}})_{(N,N')}&=&{1\over 2^D}\sum_{\vec{\sigma}}
(-1)^{\vec{\varepsilon}\cdot\vec{\sigma}}(\nabla_{\mu}^{\vec{\sigma}})_{(N,N')}, 
\label{the gen diff op}
\end{eqnarray}
with
\begin{eqnarray}
(\nabla_{\mu}^{\vec{\sigma}})_{(N,N')}&=&\left\{
\begin{array}{l}
{1\over 2a}\left(\delta_{N,N'}U_{2N+\vec{\sigma},\mu}-
\delta_{N-\hat\mu,N'}U^{\dag}_{2N+\vec{\sigma}-\hat\mu,\mu}\right),
\qquad\sigma_{\mu}=0,\\[18pt]
{1\over 2a}\left(\delta_{N+\hat\mu,N'}U_{2N+\vec{\sigma},\mu}-
\delta_{N,N'}U^{\dag}_{2N+\vec{\sigma}-\hat\mu,\mu}\right),
\qquad\sigma_{\mu}=1.
\end{array}
\right.
\label{nabla}
\end{eqnarray}
Note that $\sigma_{\mu}=0$ and $1$ give the backward and
forward difference operators respectively. 
As observed in eq. (\ref{the gen diff op}), 
the $D$-dimensional bit-valued vector $\vec{\sigma}$ is dual to $\vec{\varepsilon}$.

In the last expression of eq. (\ref{(r,N) Dirac op}), the Dirac operator
is decomposed into the difference operator acting on the microcell
coordinate $(N_{\mu})$ and the matrix acting on the relative coordinate
$(r_{\mu})$.  Accordingly, we collectively treat the fermion variables
on sites in a microcell by introducing the new variables:
\begin{equation}
\Psi_r(N) \equiv \xi_{2N+r},~~~{\bar \Psi}_r(N) \equiv {\bar \xi}_{2N+r}.
\label{Psi and xi}
\end{equation}
The new fermion variables $\Psi_r(N)$ and ${\bar \Psi}_r(N)$ belong to
spinor representations of $SO(2D)$.

\section{Symmetries of Staggered Fermion}

The single component staggered fermion action is invariant under the
discrete rotation, the chiral transformation, parity and charge
conjugation.  Here we explain these symmetries.  In particular, we will
find that the $SO(2D)$ Clifford algebra plays a vital role to describe
the rotational discrete symmetry.  Also, the well-known chiral
transformation will be identified with the $U(1)_A$ chiral
transformation associated with the $SO(2D)$ Clifford algebra.

\subsection{Rotation}

Let us see how the coordinates are transformed under the $\pi/2$
rotation around the center of the microcell.  Without losing the
generality, we may take the microcell attached to the origin and rotate the
system around the center of the microcell.  We consider the $\mu(<)\nu$
rotation.  It is easy to find out the following transformation rules:
For the relative coordinates,
\begin{eqnarray}
(r_R)_{\rho}= \left\{\begin{array}{ll}
           r_{\rho},  &   \rho \ne \mu,\nu, \\
           1 - r_{\nu},   &   \rho = \mu, \\
            r_{\mu},   &   \rho = \nu  ;
              \end{array}\right.
\end{eqnarray}
and for the coordinate $N_{\rho}$,
\begin{eqnarray}
(N_R)_{\rho} =  \left\{\begin{array}{ll}
           N_{\rho},  &   \rho \ne \mu,\nu, \\
           -   N_{\nu},  &   \rho = \mu, \\
             N_{\mu},  &   \rho = \nu    . 
              \end{array}\right.
\end{eqnarray}
Here the subscript $R$ implies rotated quantities.  For the simplicity
of the notation, we do not write the subscripts $\mu \nu$ explicitly.

The transformation of the generalized difference operator is found to be
\begin{eqnarray}
D^R_{\rho,\vec{\varepsilon}}(N,N')
=  \left\{\begin{array}{ll}
 (-1)^{\varepsilon_{\mu}}D_{\rho, \vec{\varepsilon}_R}(N_R, N'_R), & 
  \rho \ne \mu,\nu, \\
           (-1)^{\varepsilon_{\mu}+1} D_{\nu, \vec{\varepsilon}_R}(N_R, N'_R),  & 
  \rho = \mu, \\
           (-1)^{\varepsilon_{\mu}}   D_{\mu, \vec{\varepsilon}_R}(N_R, N'_R), &
   \rho = \nu, 
              \end{array}\right.
\label{rotation of D}
\end{eqnarray}
where
\begin{eqnarray}
(\varepsilon_{R})_{\rho} = \left\{\begin{array}{ll}
\varepsilon_{\rho}, &  \rho \ne \mu,\nu, \\
           \varepsilon_{\nu},   &   \rho = \mu, \\
           \varepsilon_{\mu},   &   \rho = \nu.
              \end{array}\right.
\end{eqnarray}

The staggered Dirac operator consists of two factors,
$D_{\rho,\vec{\varepsilon}}$ and $\Gamma_{\rho,\vec{\varepsilon}}$.
Having obtained the transformation property of the former, we need to
find the transformation of the latter that compensates the sign factors
in eq.~(\ref{rotation of D})  so that the Dirac operator transforms as
\begin{eqnarray}
D_{\rm st}^R=V_{\mu\nu} D_{\rm st} V_{\mu\nu}^{\dagger}
\label{tf Dst}
\end{eqnarray}
with a matrix $V_{\mu\nu}$.  When this is realized, the
transformations of $\Psi$ and ${\bar \Psi}$ are to be
\begin{eqnarray}
\Psi_{R }(N) &=& V_{\mu\nu} \Psi(N_R ), \nonumber\\
{\bar{\Psi}}_R(N)  &=& \bar{\Psi}(N_{R} ) V_{\mu\nu}^{\dagger},
\end{eqnarray}
in order for the single staggered fermion action, 
\begin{equation}
 S= \sum {\bar \Psi} D_{st} \Psi,
\label{the action}
\end{equation}
to be invariant.

In order to realize (\ref{tf Dst}), we require that
$\Gamma_{\rho,\vec{\varepsilon}}$ transform as
\begin{eqnarray}
V_{\mu\nu} \Gamma_{\rho,\vec{\varepsilon}}V_{\mu\nu}^{\dagger}= 
 \left\{\begin{array}{ll}
        (-1)^{\varepsilon_{\mu}}\Gamma_{\rho, \vec{\varepsilon}_R},   &  
 \rho \ne \mu,\nu, \\
       (-1)^{\varepsilon_{\mu}+1} \Gamma_{\nu, \vec{\varepsilon}_R},  & 
  \rho = \mu, \\
       (-1)^{\varepsilon_{\mu}}   \Gamma_{\mu, \vec{\varepsilon}_R}, &
   \rho = \nu. 
              \end{array}\right.
\label{tf of Gamma}
\end{eqnarray}
The matrix $V_{\mu\nu}$ is to found from this condition (\ref{tf of
Gamma}).  After some efforts, we find the matrix of desired property,
\begin{eqnarray}
 V_{\mu\nu} = \frac{1}{2}\Gamma_{2D+1}(\tilde{\gamma}_{\mu}- \tilde{\gamma}_{\nu})
(1+\gamma_{\mu}\gamma_{\nu}).
\label{Vmunu}
\end{eqnarray}
Note that the condition
(\ref{tf of Gamma}) determines the matrix $V_{\mu\nu}$ up to a phase
factor.

By combining eqs.~(\ref{rotation of D}) and (\ref{tf of Gamma}), we
easily see that the Dirac operator transforms as eq.~(\ref{tf Dst}) .
Therefore, we conclude that the single staggered fermion action is
invariant under the discrete rotational transformation.

{}From eq.~(\ref{tf of Gamma}), it is easy to find the action of
$V_{\mu\nu}$ on the basis of the $SO(2D)$ Clifford algebra:
\begin{eqnarray}
V_{\mu\nu} \gamma_{\rho}V_{\mu\nu}^{\dagger}= 
 \left\{\begin{array}{ll}
           \gamma_{\rho},  &   \rho \ne \mu,\nu, \\
             - \gamma_{\nu},  &   \rho = \mu, \\
             \gamma_{\mu},  &   \rho = \nu, 
              \end{array}\right.
\qquad
V_{\mu\nu} \tilde{\gamma}_{\rho}V_{\mu\nu}^{\dagger}= 
 \left\{\begin{array}{ll}
           \tilde{\gamma}_{\rho},  &   \rho \ne \mu,\nu, \\
             \tilde{\gamma}_{\nu},  &   \rho = \mu, \\
             \tilde{\gamma}_{\mu},  &   \rho = \nu. 
              \end{array}\right.
\label{action of V}
\end{eqnarray}
Here, note that ${\tilde \gamma}_{\mu}$ do not
transform as a vector: The action of $V_{\mu\nu}$ simply {\it
exchanges} ${\tilde \gamma}_{\mu}$ for ${\tilde \gamma}_{\nu}$.  

Related to the transformation in eq.~(\ref{action of V}), there is a
further curious feature of the matrix $V_{\mu\nu}$.  We studied the
$\pi/2$ rotation around the center of a microcell which leaves the system
invariant.  This intuitively obvious rotation causes a quite nontrivial
transformation among sites in a microcell, or the weight lattice of the
$SO(2D)$ spinor representation.  We found that this was achieved by the
matrix $V_{\mu\nu}$.  In eq.~(\ref{action of V}), note that negative is
the determinant of the transformation matrix realized on the matrices
$\gamma_{\mu}$ and ${\tilde \gamma}_{\mu}$.  Therefore the
transformation generated on the $2D$ gamma matrices belongs to $O(2D)$ but
not to $SO(2D)$.  In the continuum limit, the staggered Dirac operator
becomes flavor blind.  As a result, the transformation in eq.~(\ref{tf
Dst}) reduces to the $\pi/2$ rotation in the $D$-dimensional space.

\subsection{Chiral symmetry}


It is known that the staggered fermion action is invariant under the transformation
\begin{eqnarray}
\xi_n^{\prime}=(-1)^{|n|}\xi_n,~~~~~{\bar \xi}_n^{\prime}=(-1)^{|n|+1}{\bar \xi}_n.
\label{standard chiral tf}
\end{eqnarray}
This symmetry forbids the fermion mass term, even for odd dimensions.
In this sense, the symmetry has been regarded as a sort of chiral
symmetry for the staggered fermion, though the origin of the symmetry
has not been clearly understood.  Here we show that this symmetry is
nothing but the $U(1)_A$ chiral transformation associated with the $SO(2D)$ Clifford
algebra,
\begin{eqnarray}
\Psi' = \exp (i\theta\Gamma_{2D+1})\Psi, ~~ \bar{\Psi}'= \bar{\Psi}
\exp (i\theta\Gamma_{2D+1}),
\label{chiral tf on Psi}
\end{eqnarray}
with
\begin{eqnarray}
\Gamma_{2D+1}(r,r')&=&
(-i)^D(-1)^{\frac{D(D-1)}{2}}(\gamma_1 \cdots \gamma_D \tilde{\gamma}_1
\cdots \tilde{\gamma}_D)(r,r')\nonumber\\ 
&=&
(\sigma_3\otimes\cdots\otimes\sigma_3)(r,r') = (-1)^{\vert r
\vert}\delta_{r,r'}.
\end{eqnarray}
Taking $\theta=\pi/2$ in eq.(\ref{chiral tf on Psi}), we have
$\Psi_r^{\prime}(N) = i (-1)^{|r|} \Psi_r (N)$ and ${\bar
\Psi}_r^{\prime}(N) = i (-1)^{|r|} {\bar \Psi}_r (N)$, that can be
rewritten as eq. (\ref{standard chiral tf}) for $\xi$ and $\bar \xi$
variables.

By using the properties,
\begin{eqnarray}
\{\Gamma_{2D+1},\gamma_{\mu}\} =
\{\Gamma_{2D+1},\tilde{\gamma}_{\mu}\} =
\{\Gamma_{2D+1},\Gamma_{\mu,\vec{\varepsilon}}\} = 0,
\label{tf of gammas}
\end{eqnarray}
we obtain the relation
\begin{eqnarray}
\exp{(i\theta\Gamma_{2D+1})}D_{\rm st} \exp{(i\theta\Gamma_{2D+1})} 
= D_{\rm st}.
\end{eqnarray}
Therefore, the action (\ref{the action}) is invariant under the chiral
transformation in any even as well as odd dimensions.  

In the rest of this section, we describe discrete symmetries starting
with the parity symmetry.

\subsection{Parity and charge conjugation}

We take the reflection as the definition of the parity.  The reflection
along $\mu$ direction, $P_{\mu}$, only affects the $\mu$-components of
coordinates,
\begin{eqnarray*}
[N_{\mu}]_{ P_{\mu}}= -   N_{\mu},~~~~~[\sigma_{\mu} ]_{ P_{\mu}}= 1 -   \sigma_{\mu}, 
\end{eqnarray*}
\noindent
and a link variable transforms as, 
\begin{eqnarray*}
[U_{2N+\vec{\sigma},\rho} ]_{ P_{\mu}}= 
 \left\{\begin{array}{ll}
           U_{[2N+\vec{\sigma}]_{P_{\mu}},\rho},  &   \rho \ne \mu, \\
         U^{\dagger}_{[2N+\vec{\sigma}]_{P_{\mu}},\rho},  &   \rho = \mu. 
              \end{array}\right.
\end{eqnarray*}
Therefore, we find
\begin{eqnarray}
[D_{\rho,\vec{\varepsilon}} ]_{ P_{\mu}}= 
 \left\{\begin{array}{ll}
           (-1)^{\varepsilon_{\mu}}D_{\rho,\vec{\varepsilon}},  &   \rho \ne \mu, \\
         (-1)^{\varepsilon_{\mu}+1}D_{\mu,\vec{\varepsilon}},  &   \rho = \mu. 
              \end{array}\right.
\label{D parity}
\end{eqnarray}
The matrix $V_{\mu}\equiv \Gamma_{2D+1}{\gamma}_{\mu}$ produces the
appropriate sign factors by acting on $\Gamma_{\rho,\varepsilon}$:
\begin{eqnarray}
V_{\mu}\Gamma_{\rho,\vec{\varepsilon}}V^{\dagger}_{\mu} = 
 \left\{\begin{array}{ll}
           (-1)^{\varepsilon_{\mu}}\Gamma_{\rho,\vec{\varepsilon}},  &   \rho \ne \mu, \\
    (-1)^{\varepsilon_{\mu}+1}\Gamma_{\mu,\vec{\varepsilon}},  &   \rho = \mu~.
              \end{array}\right.
\label{Gamma parity}
\end{eqnarray}
{}From eqs.~(\ref{D parity}) and (\ref{Gamma parity}), we find the
Dirac operator transforms as $[D_{\rm st}]_{P_{\mu}}=V_{\mu}D_{\rm
st}V_{\mu}^{\dagger}$.  This implies the action invariance under the
parity.
In particular, under the simultaneous reflections of all the
directions, coordinates change their signs, $[n_{\mu}]_{PT}= -
n_{\mu}$, and the Dirac operator is transformed as $[D_{\rm
st}]_{PT}=\gamma_1 \cdots \gamma_{D}D_{\rm st}\gamma_{D} \cdots
\gamma_{1}$.

The charge conjugation invariance of the action can be expressed as the
following condition on the Dirac operator,
\begin{eqnarray}
C^{-1}D_{\rm st}(N, N')C = D_{\rm st}(N', N).
\label{charge conjugation}
\end{eqnarray}

First we note that the generalized difference operator,
$D_{\mu,\vec{\varepsilon}} (N,N')$, produces a sign factor under the
exchange of $N$ and $N'$:
\begin{eqnarray*}
D_{\mu,\vec{\varepsilon}}(N',N) = (-1)^{\varepsilon_{\mu}+1}
D_{\mu,\vec{\varepsilon}}(N,N').
\end{eqnarray*}
This is due to the property of $\nabla_{\mu}^{\vec{\sigma}}$ given in
eq.~(\ref{nabla}) under the exchange of $N$ and $N'$,
\begin{eqnarray*}
(\nabla_{\mu}^{\vec{\sigma}})_{(N',N)} = (-1)^{{\sigma}_{\mu}+1}
(\nabla_{\mu}^{(1-2\sigma_{\mu}) \vec{e}_{\mu}+\vec{\sigma}})_{(N,N')}.
\end{eqnarray*}

To obtain eq.~(\ref{charge conjugation}),
$\Gamma_{{\mu},\vec{\varepsilon}}$ has to produce the same sign factor
under the action of the charge conjugation matrix
\begin{eqnarray*}
C^{-1}\Gamma_{{\mu},\vec{\varepsilon}} C = 
(-1)^{{\varepsilon}_{\mu}+1}\Gamma_{{\mu},\vec{\varepsilon}}^t
= -\Gamma_{\mu,\vec{\varepsilon}}.
\label{C on Gamma}
\end{eqnarray*}
This determines how the charge conjugation matrix $C$ acts on
$\gamma_{\mu}$ and ${\tilde \gamma}_{\mu}$:
\begin{eqnarray}
C^{-1} \gamma_{\mu} C &=& \eta'\gamma_{\mu}^t = \eta'\gamma_{\mu},\nonumber\\
C^{-1} \tilde{\gamma}_{\mu} C &=& -\eta'\tilde{\gamma}_{\mu}^t = 
\eta'\tilde{\gamma}_{\mu}
\label{cond on C}
\end{eqnarray}
with $\eta'=-1$.  Therefore our charge conjugation matrix is
$C=\Gamma_{2D+1}$.\footnote{The matrix $C=1$ satisfies the
condition (\ref{cond on C}) with $\eta'=1$.}

By combing all the results stated above, we now know that the action
of the staggered fermion is invariant under the lattice version of the
CPT transformation.

As we have already mentioned, the symmetries of the staggered fermion
have been studied earlier in Ref.~\cite{Doel-Smit:1983dg}.  The authors
refer to the symmetries as shift symmetry (the translation by twice of the
lattice constant), rotational symmetry, axis reversal (parity)
symmetry, interchange (charge conjugation) symmetry, $U(1)_V$ (fermion
number) symmetry and $U(1)_A$ (chiral) symmetry.  As the spatial
rotation, they chose a site as the center of the rotation,
while we did the center of a microcell~(hypercube)
\cite{Itoh:2001rx,Itoh:2002nq}.  We have chosen the center to respect
the geometrical structure of the staggered fermion.  It may be interesting
to realize that two rotations are related by a dual transformation:
The center of a microcell is a site on the dual lattice.  Another major
difference of our approach from that of Ref.~\cite{Doel-Smit:1983dg}
is that we utilized the $SO(2D)$ Clifford algebra to express all the
symmetries.

\section{Scalar Operators} 

The results we have obtained help us to write down operators
with definite rotational and discrete symmetry properties.  Let us
consider the scalar mesons that can be constructed with ``local
variables'', in the sense that all the fields are associated with a
single microcell.  We find four such operators as given
bellow:
\begin{eqnarray}
M_1(N)  & = &  \sum_r \bar{\Psi}_r(N) \Psi_r(N),   \nonumber \\
M_2(N)  & = & \sum_{r,r'} \bar{\Psi}_r(N) (\Gamma_{D+1})_{r,r'} \Psi_{r'}(N),   
  \nonumber \\
M_3(N) & = & \sum_{\mu,r,r'} \bar{\Psi}_r(N) (\tilde{\gamma}_{\mu})_{r,r'} \Psi_{r'}(N) ,   \label{mesons} \\
M_4(N) & = &  \sum_{\mu,r,r'} \bar{\Psi}_r(N)
(\Gamma_{D+1}\tilde{\gamma}_{\mu})_{r,r'} \Psi_{r'}(N) ,  
\nonumber
\end{eqnarray}
\noindent
where
\begin{eqnarray*}
\Gamma_{D+1} &=& \sigma_2\otimes \sigma_1 \cdots \sigma_2\otimes\sigma_1  
= (-i)^{\frac{D}{2}} \gamma_1 \cdots \gamma_D
\end{eqnarray*}
\noindent
for $D=$ even and
\begin{eqnarray*}
\Gamma_{D+1} &=& \sigma_1\otimes \sigma_2 \cdots \otimes\sigma_1 
= (-i)^{\frac{D-1}{2}} \gamma_1 \cdots \gamma_D
\end{eqnarray*}
for $D=$ odd.  The presence of $\Gamma_{D+1}$ for odd dimensions is
due to the fact that the spinor representation is reducible as a
representation of the $SO(D)$ algebra. The operators $M_3$ and $M_4$
are scalars since ${\tilde \gamma}_{\mu}$ are simply exchanged under
rotations, as we have observed earlier in eq.~(\ref{action of V}).

It is quite important to realize that only $M_1$ and $M_2$ are
considered in studies of the staggered fermion.  Though the
operators $M_1$ and $M_3$ (or $M_2$ and $M_4$) are distinguished by
the chiral symmetry, they mix up due to the finite mass term.  So we
have to resolve this mixing before taking the chiral limit.

\section{Discussions} 

We studied exact symmetries of the staggered fermion that are present
even before taking the continuum limit.  The sites on the
$D$-dimensional hypercube, the microcell, form the weight lattice of the
spinor representation of $SO(2D)$.  That is the very reason why we
obtain the spinor representation of the rotational group $SO(D)$.  We
described properties of the Dirac operator under the discrete rotation
and showed the invariance of the action.  We also defined the lattice
versions of $P$, $PT$ and $C$ transformations and showed the action
invariance.  It is worth pointing out that field transformations under
the rotation and discrete transformations are uniquely defined since the
fermion variables are in an irreducible representation of the $SO(2D)$
Clifford algebra.  {}From our formulation based on the $SO(2D)$ Clifford
algebra, we obtained the following results. 1) The transformation in
eq.~(\ref{standard chiral tf}), that had been regarded as a sort of
chiral symmetry, is now identified as the chiral symmetry associated
with the $SO(2D)$ Clifford algebra. 2) Based on properties of fields
under exact symmetries, we constructed meson operators, including two
new operators, one scalar and one pseudoscalar operators.  This fact
implies operator mixings.  So we have to take account of mixings to
reach a reliable result in simulations.

The staggered fermion, by its construction, has its spinor and flavor
degrees of freedom in the geometrical structure.  In the $\xi$
variables, the action is invariant under the site-wise gauge
transformation.  After rewriting it in terms of the $\Psi$ variables
defined in eq.~(\ref{Psi and xi}), we still have the same gauge
symmetry.  However, a part of the gauge symmetry distinguishes the
components of $\Psi$, the spinor and/or flavor indices.  This looks
peculiar and we expect this part of the gauge symmetry is
absent in the proper continuum limit.  The question how it happens
certainly deserves a further study.

The results reported here could be useful for realizing supersymmetry on
lattice.  In our works\cite{Itoh:2001rx,Itoh:2002nq} aiming at a
realization of lattice supersymmetry, we presented models that possess a
Grassmannian symmetry relating bosonic and fermionic variables.  The
geometrical structure of the staggered fermion, explained in the present
paper, played a crucial role to realize the symmetry.  Since we obtained
a susy-like transformation for the fermion in the naive continuum limit,
we expect that the Grassmannian symmetry is related to some
supersymmetry in an appropriate continuum limit.

As we used the staggered fermion in \cite{Itoh:2001rx,Itoh:2002nq},
the expected supersymmetry may be an extended supersymmetry.  If it is
so, scalar degrees of freedom must be hidden in the bosonic
variables.  Therefore it is very important to fully understand
rotational properties of the theory including the link variables.

\acknowledgments

\noindent
This work is supported in part by the Grants-in-Aid for Scientific
Research No. 13135209, 15540262, 16340067 from the Japan Society for the
Promotion of Science.  The authors thank the Yukawa Institute for
Theoretical Physics at Kyoto University.  Discussions during the YITP
workshop YITP-W-04-08 on ``Summer Institute 2004'' were useful to
complete this work.


\begin{thebibliography}{99}

\bibitem{Kogut:1974ag} J.~B.~Kogut and L.~Susskind, \prd{11}{1975}{395}

\bibitem{Susskind:1976jm} L.~Susskind, \prd{16}{1977}{3031}

\bibitem{Kawamoto:1981hw} N.~Kawamoto and J.~Smit, \npb{192}{1981}{100}

\bibitem{Kluberg-Stern:1983dg} F.~Gliozzi, \npb{204}{1982}{419};\\
H.~Kluberg-Stern, A.~Morel, O.~Napoly and B.~Petersson, \npb{220}{1983}{447}

\bibitem{Doel-Smit:1983dg} C.~P.~van den Doel and J.~Smit, \npb{228}{1983}{122};\\
M.~F.~L.~Golterman and J.~Smit, \npb{245}{1984}{61}

\bibitem{Itoh:2001rx} K.~Itoh, M.~Kato, H.~Sawanaka, H.~So and N.~Ukita,
	\ptp{108}{2002}{363} [hep-lat/0112052]

\bibitem{Itoh:2002nq} K.~Itoh, M.~Kato, H.~Sawanaka, H.~So and N.~Ukita,
	\jhep{0302}{2003}{033} [hep-lat/0210049]
\end{thebibliography}
\end{document}